# Intelligent Nano-Fingerprinting: An Efficient and Precise Approach for Liquid Biopsy

**Authors:** Yuxin Yang[1], Hexiang Bai[2], Meihua Shangguan[3,4,5], Siying Shao[1], Yizheng Fang[1], Long Yi[1], Haozhong Ma[6], Hongxia Xu[7], Xiawei Li[3,4,5,8], Yulian Wu[3,4,5,9], Zhenrong Zheng[1], Xu Liu[1], Jian Wu[10,11*], Longhua Tang[1*]

**Affiliations:**

[1] State Key Laboratory of Extreme Photonics and Instrumentation, College of Optical Science and Engineering, Zhejiang University, Hangzhou 310027, China

[2] School of Public Health, Zhejiang University School of Medicine, Hangzhou 310058, China

[3] Second Affiliated Hospital, Zhejiang University School of Medicine, Hangzhou, Zhejiang, China

[4] Key Laboratory of Cancer Prevention and Intervention, China National Ministry of Education, Cancer Institute, Second Affiliated Hospital, Zhejiang University School of Medicine, Hangzhou, Zhejiang, China

[5] Cancer Center, Zhejiang University, Hangzhou, Zhejiang, China

[6] Department of Pathology & Pathophysiology, Zhejiang University School of Medicine, Hangzhou 310058, China

[7] WeDoctor Cloud and Liangzhu Laboratory, Hangzhou, 310000, China

[8] Harvard T.H. Chan School of Public Health, Harvard University, Boston, MA, USA

[9] Department of Surgery, Fourth Affiliated Hospital, Zhejiang University School of Medicine, Yiwu, Zhejiang, China

[10] State Key Laboratory of Transvascular Implantation Devices and TIDRI, Hangzhou, 310009, China

[11] Zhejiang Key Laboratory of Medical Imaging Artificial Intelligence, Hangzhou 310058, China

* Corresponding authors: wujian2000@zju.edu.cn (J.W.); lhtang@zju.edu.cn (L.T.)

**Abstract**

Biological matrices are rich in information related to life processes, serving as invaluable media for assessing an individual's overall physiological status and its dynamic fluctuations, as well as crucial foundations for disease diagnosis. However, the inherent complexity of these matrices, coupled with our incomplete understanding of their full composition, presents significant challenges for comprehensive analysis and accurate diagnostic interpretation. The advent of single-molecule technologies has revolutionized biomedical research, enabling the direct observation of life processes at the molecular scale. We have proposed an Intelligent Nano-Fingerprinting strategy based on single-molecule nanopore technology, designed to capture the global molecular fingerprints of complex plasma matrices. Furthermore, we developed an intelligent algorithmic model capable of achieving precise classification of plasma samples. This approach is characterized by its simplicity, efficiency, and considerable potential for large-scale adoption and transferable applications.

**Main**

Liquid biopsy represents an advanced research direction in clinical pathology.[1–3] Approximately 60% of an adult's body weight consists of fluid matrices, which are essential for sustaining normal life activities.[4] Among these, blood holds a central position in all bodily fluids. This is not only because every cell releases products of its physiological processes into the bloodstream, but also because blood continuously exchanges with the tissue microenvironments throughout the body via the circulatory system, thereby serving as a comprehensive reflection of the body's macroscopic physiological state.[5,6]

By detecting molecular biomarkers in blood, associations with specific diseases can be established, enabling diagnosis. For example, PCR[7] and ELISA[8] are used to detect nucleic acid and protein biomarkers, respectively. Relying solely on a single biomarker often suffers from insufficient specificity, while combining multiple biomarkers can improve diagnostic accuracy. Furthermore, multi-omics technologies provide a comprehensive assessment of blood components by integrating multi-dimensional molecular information (including genomics, transcriptomics, proteomics, and metabolomics).[9,10] However, such techniques involve highly complex sample preprocessing steps, requiring separate extraction and enrichment of different types of components. This process consumes significant sample volumes, risks introducing additional interference if not handled properly, and presents operational challenges due to the difficulty in standardizing workflows for different component types.[11,12] Given that the molecular composition of blood forms a complex network, even seemingly simple physiological changes may correlate with variations in multiple types of molecules. Additionally, specific interactions may exist among different molecules, and additional separation steps could lead to loss of information.[5,13] From a holistic perspective, in situ reflection of molecular-level differences in blood may be critical for precision medicine.

We propose an Intelligent Nano-Fingerprinting approach, based on single-molecule nanopore technology, which enables in situ detection of the overall molecular heterogeneity of blood in a label-free and amplification-free manner. The sample requires only dilution without complex preprocessing for nanopore detection, with each test consuming only microliter-level volumes. By incorporating artificial intelligence analytical algorithms, the nanopore fingerprints can be accurately mapped to sample types with high precision. This method is simple, efficient, and holds promise as an economical and reliable auxiliary tool for large-scale clinical screening.

It may facilitate the rational allocation of medical resources and, when combined with other diagnostic technologies, help elucidate disease mechanisms.

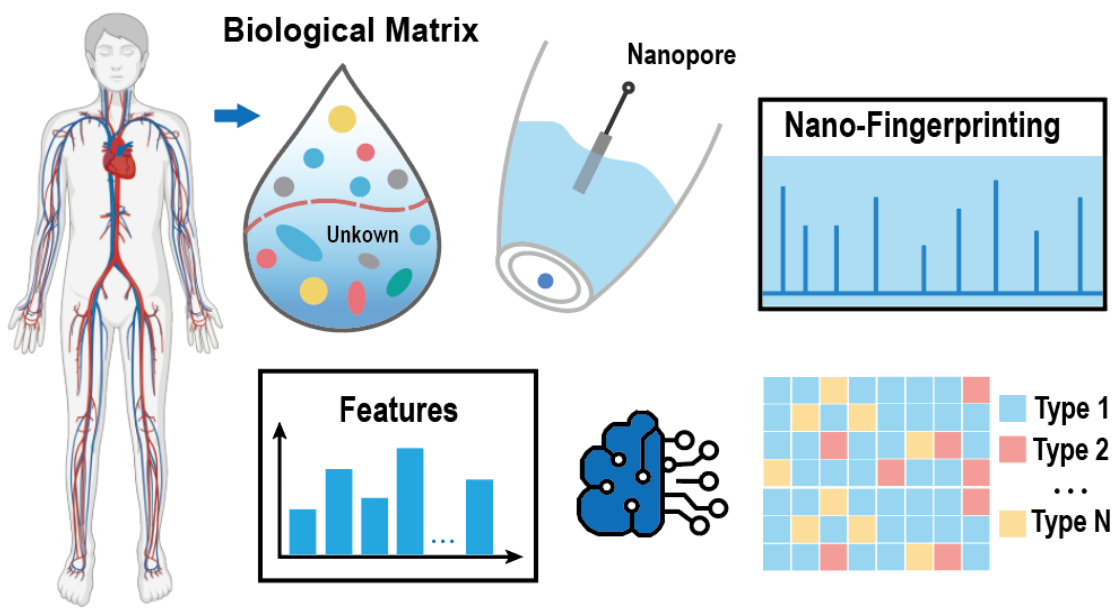

Figure 1. Intelligent Nano-Fingerprinting for liquid biopsy.

**Intelligent Nano-Fingerprinting of Plasma**

The core component of nanopore technology is a nanoscale pore that operates within an electrolyte solution. When a voltage is applied across the pore, electrolyte ions migrate directionally through it, generating a baseline current. After introducing the target analyte, molecules are driven through the nanopore under the combined effects of free diffusion, electroosmotic flow, and electrophoretic force, thereby perturbing the ionic current passing through the pore.[14,15] Variations in the ionic current reflect multiple molecular properties, including three-dimensional conformation, effective volume, and surface charge distribution. Furthermore, the rapid dynamic behavior of molecules inside the pore and their interactions with the pore wall induce transient fluctuations in the ionic current, which are typically superimposed on the background noise.[16–19] Originally developed for sequencing applications, nanopore technology has significantly advanced genomic research and found preliminary clinical applications in rapid pathogen detection and cancer diagnosis[20]. Beyond nucleic acids, the technology has been successfully applied to protein analysis[21], disease biomarker detection[22], and even small molecule analysis, including metabolites, hormones, and pharmaceuticals[23], demonstrating considerable clinical potential.

We obtained Nano-Fingerprinting of plasma through nanopore technology. The nanopores were fabricated from quartz capillaries using a laser-assisted pulling technique, which involves two critical steps: melting the quartz via laser heating and applying tensile forces to both ends to obtain two symmetrical tapered quartz capillaries, with the nanopores located at the conical tips of the capillaries (as shown in Figure 2). By adjusting parameters such as heating temperature, heating duration, tensile force magnitude, and pulling speed, the diameter of the resulting nanopores can be precisely controlled. We employed nanopores with a diameter distribution ranging from 30 to 40 nm, which yielded relatively stable and high-quality signals during plasma testing (as discussed in Supplementary Information, Section 3). Plasma samples were collected from medically individuals and diluted 10-fold in a TE buffer system (1M KCl, 10mM Tris HCl, 1mM EDTA, pH=7.8-8.2). The diluted plasma solution was injected into a microfluidic channel, and electrodes were placed in both the tapered glass capillary and the microfluidic channel, respectively. Bias voltages of 100 mV and -100 mV were applied sequentially, each lasting for 5 minutes. The typical Nano-Fingerprinting was displayed in Figure 2, which reflects molecular-level fluctuations in the plasma matrix.

Nano-Fingerprinting of different plasma samples exhibits variations (as shown in Figure S6 of the Supplementary Information), which are difficult to quantify through direct visual inspection. To enable robust comparison, we developed an intelligent algorithm for qualitative analysis. Raw fingerprinting data of plasma were first processed via wavelet transformation to correct baseline drift, followed by Savitzky–Golay (SG) filtering for noise reduction. Subsequently, feature extraction was performed across multiple dimensions to comprehensively characterize the differences among fingerprint profiles. In the time domain, major signal peaks were identified by setting thresholds, and the overall distribution of current was characterized using statistical moments. Frequency-domain features were obtained via Fast Fourier Transform (FFT) to decompose the signal into its constituent spectral components. To capture transient dynamics and non-stationary patterns, time-frequency representations were analyzed. Additionally, nonlinear metrics were computed to reflect the complex, dynamic behavior inherent in the fingerprints. These extracted features were then integrated and fed into a Gradient Boosting Decision Tree (GBDT) classifier. The model performed ensemble voting across molecular fingerprint categories, with final class assignments determined based on the highest predicted probability. This approach effectively establishes a reliable mapping from nano-fingerprinting signatures to differentiated plasma classifications.

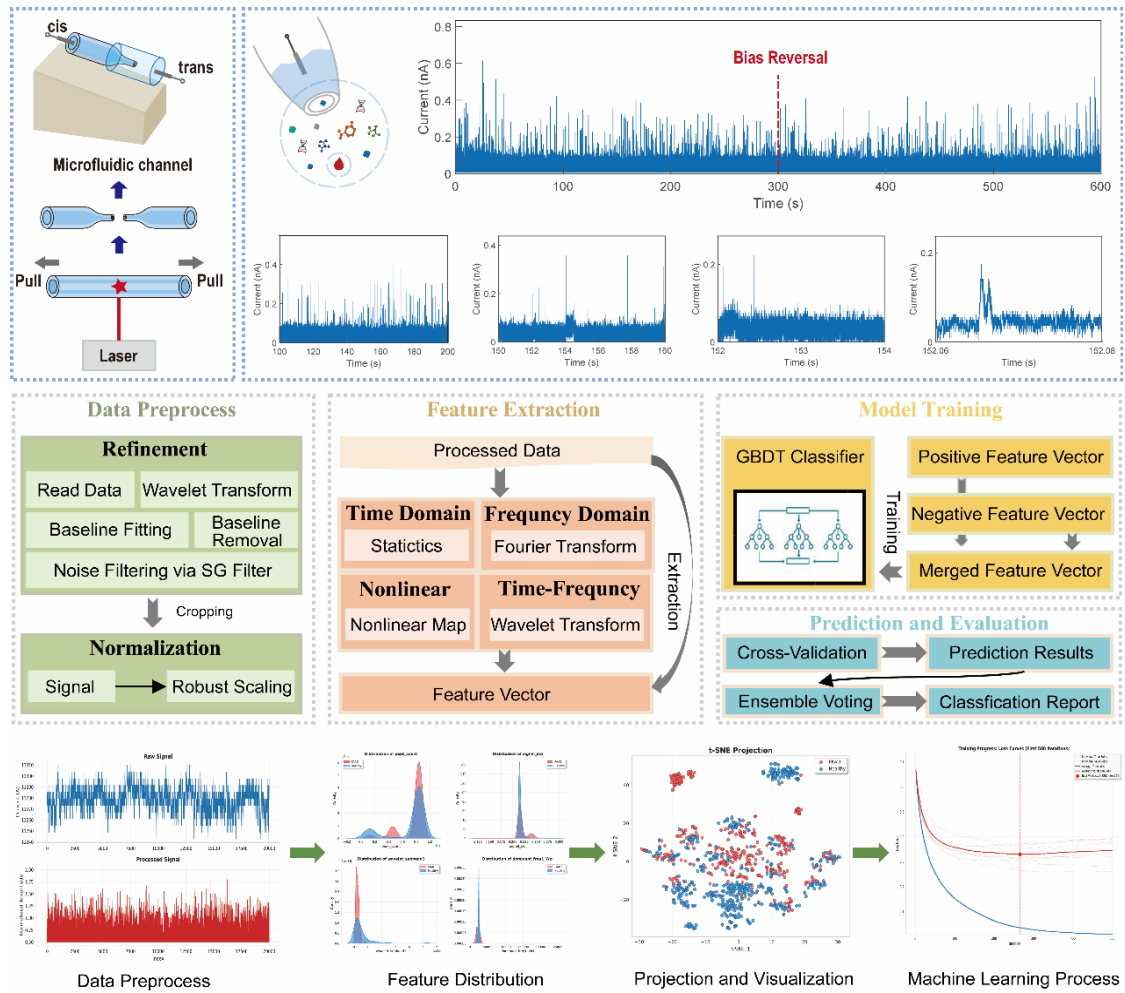

Figure 2. System Design of Intelligent Nano-Fingerprinting.

Due to fluctuations in laser power, subtle variations in nanopore morphology exist across different fabrication batches. To quantify these differences, we modeled the nanopore as a truncated hollow cone, as illustrated in Figure 3(a), and its resistive properties can be expressed as:[24]

$$R_p \approx \frac{1}{\kappa \pi r_i \tan \beta} + \frac{1}{4\kappa r_i}$$

where $R_p$ is the nanopipette resistance, $r_i$ the inner pipette radius, $\kappa$ is the solution conductivity, and $\beta$ is the inner nanopipette half cone angle. SEM measurements revealed that the variation in $\beta$ among different nanopore devices is minimal (as shown in Figure 3(a) and Table S1 of Supplementary Information). Therefore, the equation can be simplified, indicating that $R_p$ is approximately inversely proportional to $r_i$. Under the same bias voltage, nanopores with different diameters produce differentiated baseline ionic currents for the same sample, and the relative current changes induced by molecule translocations also vary. To eliminate this device-induced measurement bias, we introduced a correction factor[25] (detailed in the Supplementary Information, Section 2). Before and after correction, the model performance showed little difference (as shown in Figure 3(b), with AUC values of 0.8217 and 0.8451 before and after correction, respectively), demonstrating its good robustness against device heterogeneity.

The use of the microfluidic channel significantly reduces the sample consumption (only 2 μL of plasma per test). The microfluidic channel is fabricated from a quartz capillary with an outer diameter of 2.5 mm and an inner diameter of 1.5 mm, yielding a volume of approximately 20 μL. By comparing with a conventional chamber, which has a volume of 200 μL (the diameter is approximately 10 mm), we found that the microfluidic channel significantly enhanced signal enrichment capability (see Figure S5 in Supplementary Information). When their collected data is used for modeling, the microfluidic channel (AUC = 0.9744) significantly outperforms that of the chamber (AUC = 0.6781), as shown in Figure 3(c). We attribute this primarily to two reasons. Firstly, microfluidic channel has a significantly smaller volume compared to the chamber. For the same sample, although the absolute quantity of biomolecules remains the same, the concentration gradient is higher, and the average distance for diffusion to the vicinity of the nanopore is significantly shortened. This greatly enhances the probability of molecular capture by the nanopore.[26,27] Secondly, the capillary force generated by the microfluidic channel promotes the accumulation of molecules along the inner glass surface, which subsequently enhances their electrokinetic delivery toward the nanopore.

We also investigated the capture of biomolecules from plasma by the nanopore under an applied electric field. In our experimental setup, the nanopore side was connected to the cis electrode, while the sample reservoir was connected to the trans electrode. Biomolecules are primarily influenced by three forces: diffusion, electroosmotic flow, and electrophoretic force. Considering the differences in surface charge among different molecules, we performed data acquisition under both positive and negative bias voltages. As shown in Figure 3(d), we observed that combining data from both positive and negative bias voltages for modeling yielded the highest performance (AUC = 0.8250), compared to models trained solely on positive or negative bias voltage data. This indicates that integrating both bias polarities contributes to a more comprehensive characterization of sample features. The influence of measurement time on the model's discriminative ability is illustrated in Figure 3(e) and (f). For the three data categories (positive bias, negative bias, and combined biases), classification models were built using cumulative data segments of increasing length (the first 1 minute, first 2 minutes, …, 5 minutes). All three groups exhibited a notable change in accuracy around the 3-minute mark, after which the accuracy stabilized with minimal further variation. Considering the testing efficiency for large-scale sample analysis, a signal acquisition duration of 5 minutes under each bias condition is deemed appropriate.

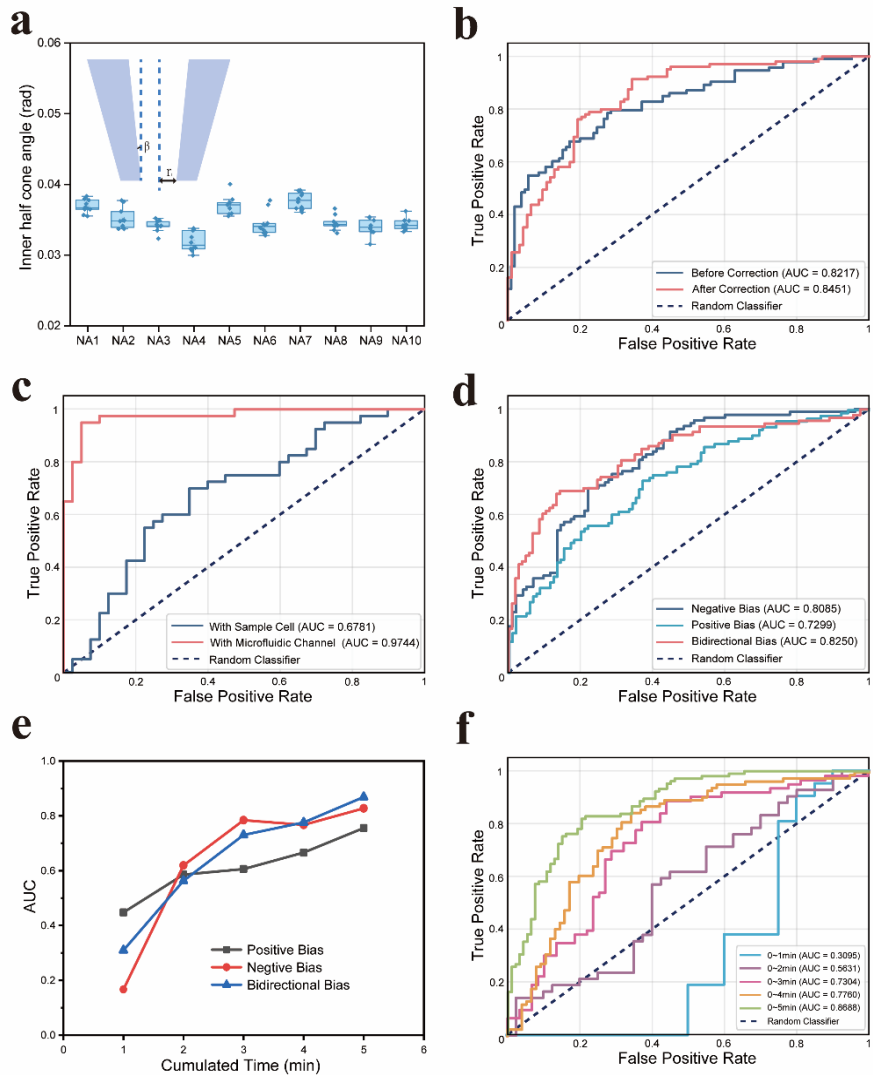

Figure 3. Detailed settings of the nanopore measurement. (a) Variation in the internal half-cone angles of 10 nanopores fabricated in the same batch, with each group measured 10 times repeatedly. (b) Impact of device heterogeneity correction on the model before and after adjustment. (c) Influence of using microfluidic channels versus conventional chambers as sample reservoirs on model construction. (d) Effect of modeling using only positive-pressure test data, only negative-pressure test data, or combined positive- and negative-pressure test data. (e) and (f) illustrate the impact of sampling duration on model performance. (e) shows the changes in AUC obtained when modeling separately with positive-pressure data, negative-pressure data, and combined positive- and negative-pressure data, as well as when using data from the first 1 min, the first 2 min, ⋯ up to the first 5 min for each group. (f) depicts the changes in ROC curves when modeling with combined positive- and negative-pressure data using the first 1 min, the first 2 min, ⋯ up to the first 5 min of data, respectively.

**Clinical Application Potential of Intelligent Nano-Fingerprinting**

The efficacy of the Intelligent Nano-Fingerprinting was validated through real clinical samples. Plasma from 75 healthy individuals was collected, and classification boundaries for physiological characteristics (including sex, age and BMI) were established. For sex, the categories were male and female. For age and BMI, the cutoffs were set at 55 years and a BMI of 24, respectively, determined based on the median values of the sample population. Subsequently, binary classification was performed for each of the three physiological traits. When classifying based on a single trait, the other two indicators were kept consistent. As a result, we obtained 12 classification outcomes, as shown in Figure 4(a). The prediction accuracy for each category reaches 0.8 or higher, indicating that the Intelligent Nano-Fingerprinting can effectively capture the commonalities of the same category and detect subtle differences between different categories with high sensitivity.

We then evaluated the capability of the Intelligent Nano-Fingerprinting in discriminating clinical cancer samples. A total of 46 breast cancer samples, 46 gastric cancer samples, and 46 healthy samples of human plasma were included. Representative molecular fingerprints for the three sample types are shown in Figure 4(b). The multiclass classification task adopted the "One vs Rest" (OvR) strategy, in which each class is compared against all other classes to form a set of binary classification problems. Figure 4(c) presents the confusion matrix, indicating the number of correct and incorrect predictions for each category. Class 0.0, Class 1.0, and Class 2.0 correspond to healthy, gastric cancer, and breast cancer samples, respectively. The area under the curve (AUC), computed from the ROC curve (Figure 4(d)), served as a key performance metric. The healthy class (Class 0.0) achieved an

AUC of 0.9889, reflecting the model's strong ability to differentiate healthy table from cancerous plasma. The AUC values for gastric cancer (Class 1.0) and breast cancer (Class 2.0) were 0.9152 and 0.9062, respectively, demonstrating the model's substantial capacity to discriminate between cancer types. The radar chart in Figure 4(e) offers a comprehensive visualization of the model's performance across six key metrics, each scoring 0.8 or higher, summarizing its potential for clinical diagnostics. Recall (sensitivity) and precision reflect the ability to identify and predict positive cases, whereas specificity and negative predictive value (NPV) measure its performance in identifying and predicting negative cases. In clinical diagnostics, these metrics indicate how well the model avoids missed diagnoses and misdiagnoses. The F1-score provides a harmonic mean of precision and recall, while balanced accuracy represents an average of recall and specificity, together offering an overview of overall model performance. Based on macro-averages, all metrics exceeded 87%, with specificity and NPV surpassing 93%, underscoring the model's strength in ruling out disease and minimizing false positives. The comparatively lower recall and F1-score were mainly attributable to the lower recall (80.43%) for Class 2.0 (breast cancer). Expanding the sample size may further improve model performance. Overall, this model offers a feasible approach for preliminary cancer screening.

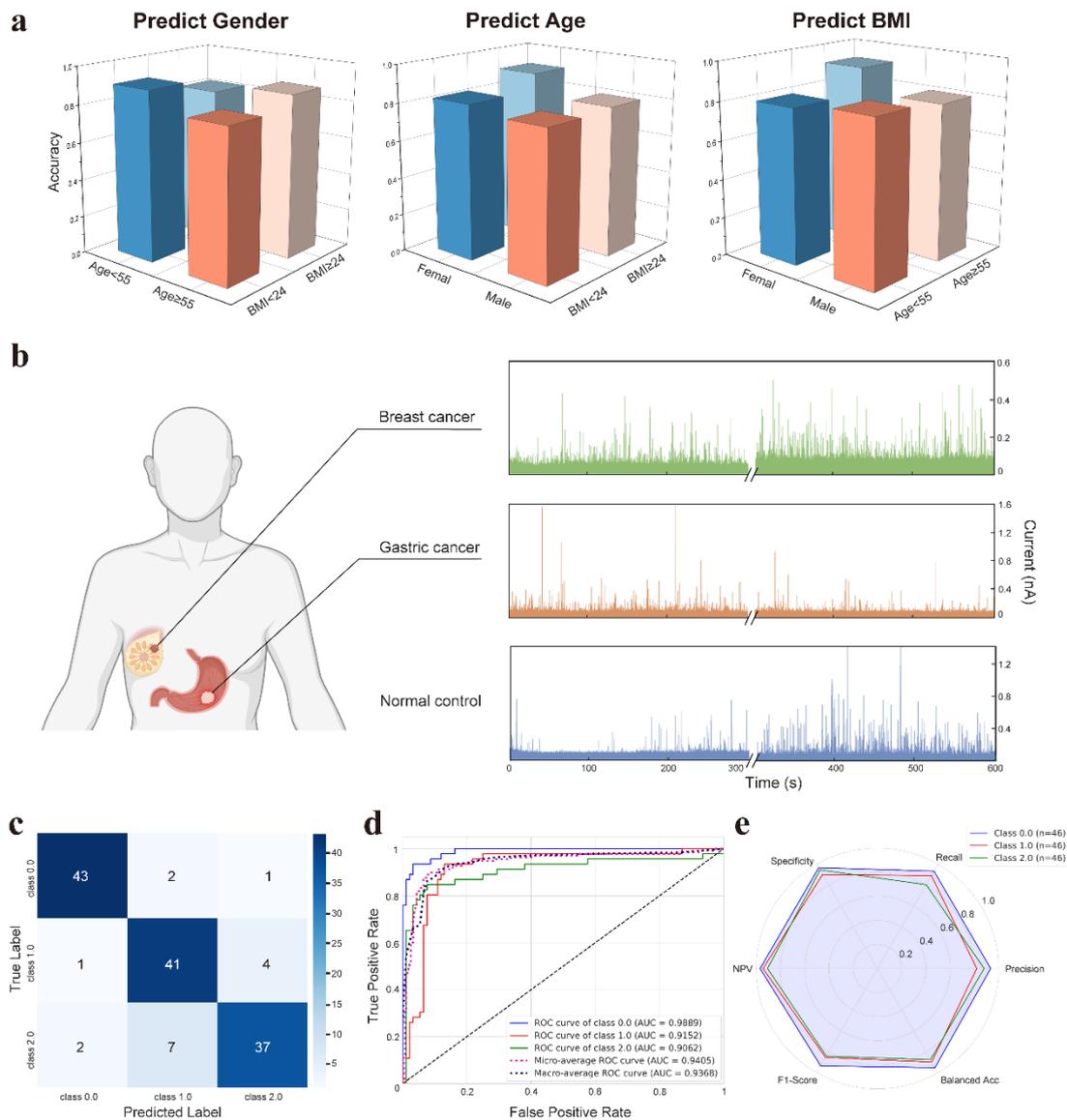

Figure 4. (a) The Intelligent Nano-Fingerprinting for classifying physiological characteristics. (b) Molecular fingerprints of breast cancer, gastric cancer, and healthy clinical human plasma samples. The (c) confusion matrix, (d) ROC curve and (e) radar chart of the model.

## Conclusions and Outlook

In summary, we have developed the Intelligent Nano-Fingerprinting and validated its potential for clinical applications through plasma analysis. The method operates without complex sample pretreatment and allows efficient acquisition of global molecular fingerprints. By leveraging multidimensional feature extraction: spanning the time, frequency, time-frequency, and nonlinear transformation domains, combined with deep learning modeling, the system achieves accurate sample classification. Our results indicate that the approach not only discriminates effectively between basic physiological traits such as sex, age, and BMI, but also delivers strong performance in detecting breast cancer and gastric cancer. The advantage of the Intelligent Nano-Fingerprinting lies in its simple operational procedure and strong generalization capability, allowing it to be easily adapted for the detection of various other biological fluid matrices. This method provides an effective tool for the diagnosis of multiple diseases and enables dynamic tracking of disease progression. In the future, conducting large-scale, multi-center clinical sample analyses and prospective studies will further enhance the robustness of the Intelligent Nano-Fingerprinting.

**Methods**

**Nanopore fabrication**

Nanopores were fabricated using a laser-assisted pipette puller (Sutter Instrument, P-2000, USA) by pulling quartz capillaries (GQF100-50-7.5, World Precision Instruments, UK). Before pulling, the capillaries (inner diameter: 0.5 mm, outer diameter: 1.0 mm, length: 7.5 cm) were treated with oxygen plasma for 30 min using a plasma cleaner (Harrick Plasma) to remove organic residues and contaminants from the quartz surface. The pulling process was performed using an optimized two-line program with the following parameters: (1) HEAT: 650; FIL: 4; VEL: 30; DEL: 170; PUL: 70, (2) HEAT: 680; FIL: 3; VEL: 20; DEL: 145; PUL: 130. The described protocol produced nanopores with an average tip size of $35 \pm 5$ nm, as confirmed by both SEM imaging and conductance measurements (see Supplementary Information, Figure S1).

**Plasma pretreatment**

Blood was collected using EDTA as an anticoagulant, followed by centrifugation at 3000 rpm for 10 minutes at room temperature. The supernatant was then collected and stored at -80°C.

**Acquisition of nanopore signals**

Electrical measurements were performed using a MultiClamp 700B amplifier and an Axon Digidata 1550B system (Molecular Devices, USA). Before being used for plasma testing, each nanopore was characterized by current-voltage (I-V) measurements in 1 M KCl buffer containing 10 mM Tris-HCl and 1 mM EDTA (pH = 8). The nanopipettes were filled with approximately 10 µL of the same buffer and inserted into a microfluidic channel prefilled with approximately 20 µL of the buffer. Two freshly prepared Ag/AgCl electrodes were placed into the nanopipette and the microfluidic channel, serving as the working and reference electrodes, respectively. Current was recorded during voltage sweeps from -400 mV to 400 mV with a step rate of 50 mV. Plasma samples were then diluted at a 1:10 ratio in the same buffer. Nanopore current signals were acquired for 5 minutes each under applied voltages of 100 mV and -100 mV.

**Ethics declarations**

This study utilized residual clinical blood samples obtained from patients diagnosed with gastric cancer (GC) or breast cancer (BC) at the Second Affiliated Hospital, Zhejiang University School of Medicine. All cancer diagnoses were pathologically confirmed. Age- and sex-matched healthy controls were selected from individuals undergoing routine physical examinations at the same institution during the same period, with no history of any malignancy. The study protocol was reviewed and approved by the Institutional Review Board (IRB) of the Second Affiliated Hospital,

Zhejiang University School of Medicine (Approval No. IR2024239). Since this study involved the analysis of anonymized, residual waste samples collected during standard clinical care procedures, and posed no additional risk to the patients, the IRB granted a waiver of informed consent.